\documentclass[10pt]{article}
\usepackage{fancyhdr}
\usepackage{extramarks}
\usepackage{amsmath}
\usepackage{amsthm}
\usepackage{amsfonts}
\usepackage{siunitx}
\usepackage{tikz}
\usepackage[plain]{algorithm}
\usepackage{algpseudocode}
\usepackage{multirow}
\usepackage{booktabs}
\usepackage{graphicx}
\usepackage{subfigure}
\usepackage[colorlinks,linkcolor=black,anchorcolor=black,citecolor=black,urlcolor=blue]{hyperref}
\usepackage{amsmath,bm}
\usepackage{booktabs}
\usepackage{mathtools}
\usepackage{amssymb}
\usepackage{caption}
\usepackage{capt-of}
\usepackage{mciteplus}
\usepackage{cite}
\usepackage{mathrsfs}
\usepackage[title,titletoc,toc]{appendix}
\usepackage{xr}
\usepackage{parskip}
\usepackage{soul}
\usepackage{textcomp}
\usepackage[colaction]{multicol}
\usepackage[switch]{lineno}
\usepackage{lipsum}
\usepackage{etoolbox}
\usepackage{longtable}
\usepackage{array}
\usepackage{tablefootnote}
\usepackage{ragged2e}
\newcolumntype{C}[1]{>{\centering\arraybackslash}p{#1}}
\usetikzlibrary{automata,positioning}
\usetikzlibrary{automata,positioning}

\begin{document}

\title{ Emerging dominant SARS-CoV-2  variants  
} 
 
\author{ Jiahui Chen$^1$, Rui Wang$^1$, Yuta Hozumi$^1$, Gengzhuo Liu$^1$, Yuchi Qiu$^1$,   Xiaoqi Wei$^1$ and \\
Guo-Wei Wei$^{1,3,4}$\footnote{
 		Corresponding author.		Email: weig@msu.edu} \\% Author name
 $^1$ Department of Mathematics, \\
 Michigan State University, MI 48824, USA.\\
 $^2$ Department of Electrical and Computer Engineering,\\
 Michigan State University, MI 48824, USA. \\
 $^3$ Department of Biochemistry and Molecular Biology,\\
 Michigan State University, MI 48824, USA. \\
 }
\date{\today} % Date for the report

\maketitle

\begin{abstract}
Accurate and reliable forecasting of emerging dominant severe acute respiratory syndrome coronavirus 2 (SARS-CoV-2) variants enables policymakers and vaccine makers to get prepared for future waves of infections.  The last three waves of  SARS-CoV-2  infections caused by dominant variants Omicron (BA.1), BA.2, and BA.4/BA.5 were accurately foretold by our artificial intelligence (AI) models built with biophysics, genotyping of viral genomes,  experimental data, algebraic topology, and deep learning. 
% We updated our AI model built with newly available viral genomes,  new deep mutational screening data, new Omicron three-dimensional (3D) structures, and new algebraic topology techniques. 
Based on newly available experimental data,  we analyzed the impacts of all possible viral spike (S) protein receptor-binding domain (RBD) mutations on the SARS-CoV-2 infectivity. 
Our analysis sheds light on viral evolutionary mechanisms, i.e., natural selection through infectivity strengthening and antibody resistance.    
We forecast that BA.2.10.4, BA.2.75, BQ.1.1, and particularly, BA.2.75+R346T, 
%BA.4.6, BF.7   
have high potential to become new dominant  variants to drive the next surge.

\end{abstract}
Keywords: COVID-19, SARS-CoV-2, Omicron,  infectivity,  subvariants, deep learning, algebraic topology. 
%\pagenumbering{roman}
%\begin{verbatim}
%\end{verbatim}
%
% {\setcounter{tocdepth}{4} \tableofcontents}
%%
\newpage
 %\clearpage
 %\pagebreak

\setcounter{page}{1}
\renewcommand{\thepage}{{\arabic{page}}}

% \begin{multicols}{2}
% \multicollinenumbers
% \linenumbers

%
 
\section{Introduction}
\begin{figure}[ht!]
	\centering
	\includegraphics[width = 0.7\textwidth]{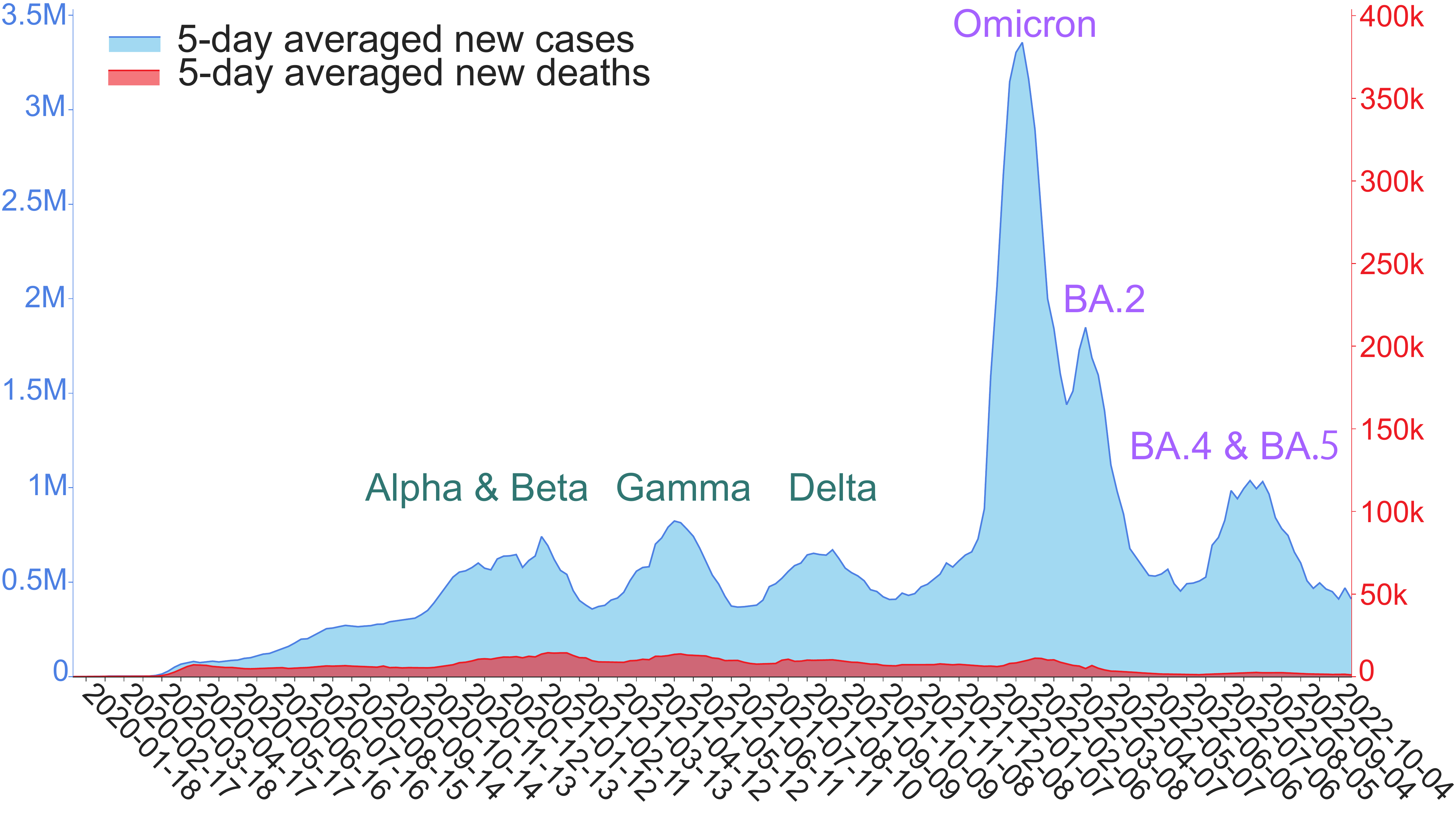}
	\caption{ Illustration of daily COVID-19 cases (light blue) and deaths (red) since 2020   \cite{Cases}. The curves are smoothed by   five-day averages. 
	}
	\label{fig:cases}
\end{figure}

In the past two years, the coronavirus disease-2019 (COVID-19) pandemic was fueled by the spread of a few dominant variants of severe acute respiratory syndrome-coronavirus-2 (SARS-CoV-2), as shown in Figure \ref{fig:cases}.  Specifically, the Alpha and Beta variants contributed to a peak of infections and deaths from October 2020 to January 2021. The Gamma variant caused another peak of infections and deaths in April and May   2021.  The Delta variant led to the third wave of COVID-19 infections and deaths around August 2021. The Omicron (B.1.1.529), which was extraordinary in its infectivity, vaccine breakthrough, and antibody resistance, created a huge spike in the world's daily infection record in December 2021 and January 2022. Omicron BA.2 subvariant rapidly replaced the original Omicron (i.e., BA.1) in March 2022. Around July 2022,  Omicron subvariants BA.4 and BA.5 took over BA.2 and became the new dominant SARS-CoV-2 variant. These variant-driven waves of infections are also associated with spikes in deaths and have given rise to tremendous economic loss.  A  life-and-death question is: what will be future dominant variants?    

Forecasting and surveillance of emerging SARS-CoV-2 variants are some of the most challenging tasks of our time. Among about half a million SARS-CoV-2/COVID-19 related publications recorded in Google Scholar, few accurately foretold the emerging SARS-CoV-2 variants. Accurate and reliable forecasting of emerging SARS-CoV-2 variants enables policymakers and vaccine makers to plan, leading to enormous social, economic, and health benefits. To foretell future variants, one must have the full understanding of the mechanisms  of viral evolution, the mechanisms of viral mutations, and the relationship between viral evolution and viral mutation.

Future variants are created through the SARS-CoV-2 evolution, in which is a  
SARS-CoV-2  evolves through changes in its RNA at molecular scale  to gain fitness over  its counterparts at the host population scale.  At the molecular scale, most mutations occur randomly. Indeed,  random genetic drift is a major mechanism of mutations, resulting in errors in various biological processes, such as replication, transcription, and translation. Additionally,  virus-virus intra-organismic recombination can alter SASR-CoV-2 genes, which has a stochastic nature too.  However, SARS-CoV-2 has a genetic proofreading mechanism facilitated by the synergistic interactions between  RNA-dependent RNA polymerase and non-structure proteins 14 (NSP14)   \cite{sevajol2014insights,ferron2018structural}. 
At the organismic scale, inter-organismic recombination happens but the resulting variants may not be clinically significant.  In contrast,  host editing of virus genes is known to be a significant mechanism for SARS-CoV-2 mutations \cite{wang2020host}. 
At the population scale, mutations occurring at molecular and organismic scales are regulated, i.e., enhanced and/or suppressed via natural selection, giving rise to SARS-CoV-2 variants with increased fitness \cite{chen2020mutations}. Therefore, natural selection is the fundamental driven force for viral evolution.   

\begin{figure}[ht!]
	\centering
	\includegraphics[width = 0.88\textwidth]{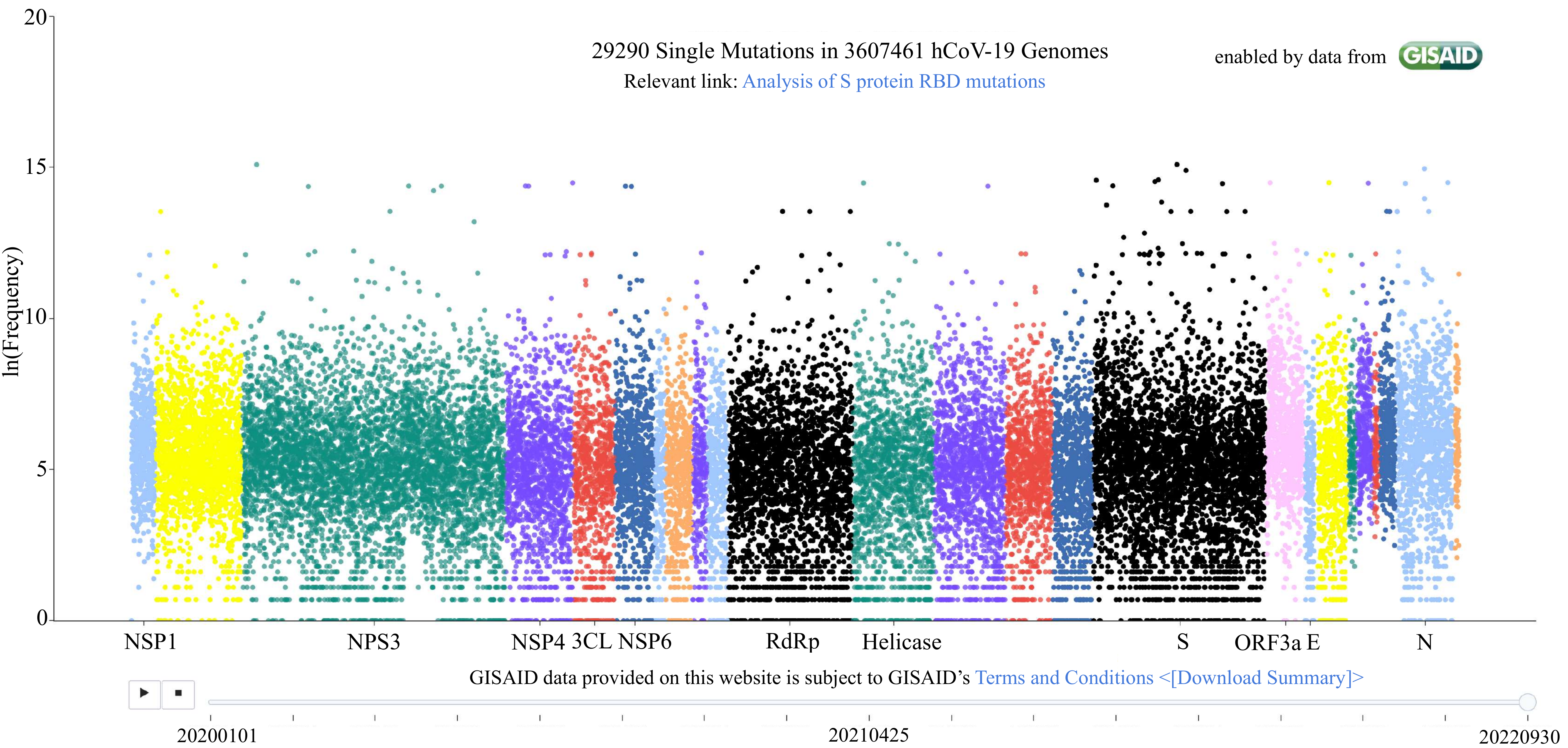}
	\caption{  Illustration of unique mutations on SARS-CoV-2 genomes extracted from patients. Each dot represents a unique mutation. The $x$-axis is the gene  position of a mutation and the $y$-axis represents its observed frequency in the natural logarithmic scale. 
	}
	\label{fig:mutationtracker}
\end{figure}

It remains to understand what controls the natural selection of SARS-CoV-2. The mechanism of SARS-CoV-2 evolution was   elusive at the beginning of the COVID-19 pandemic. Indeed,  the life cycle of SARS-CoV-2 is extremely sophisticated, involving the viral entry of host cells, the release of the viral genome, the synthesis of viral NSPs,  RNA replication, the transcription, translation, and synthesis of viral structural proteins, and the packing, assembly, and release of new viruses \cite{trougakos2021insights}. The SARS-CoV-2 mutations occur nearly randomly on all of its 29 genes, as shown in Figure \ref{fig:mutationtracker}.  Nonetheless, in early 2020, we hypothesized that SARS-CoV-2 natural selection is controlled through infectivity-strengthening mutations \cite{chen2020mutations}, which primarily occur at the viral spike (S) protein receptor-binding domain (RBD) that binds with host angiotensin-converting enzyme 2 (ACE2) to facilitate the viral cell entry  \cite{li2005bats,qu2005identification,song2005cross,hoffmann2020sars,walls2020structure}.  Our hypothesis was initially supported by  our genotyping of 15,140 SARS-CoV-2 genomes extracted from patients.  We demonstrated that among 89 unique RBD mutations, the observed frequencies of infectivity-strengthening mutations outpace those of infectivity-weakening ones in their time evolution.  Our infectivity-strengthening mechanism of natural selection was proven beyond doubt in  April 2021, with  506,768 SARS-CoV-2 genomes isolated from patients \cite{wang2021vaccine}.

However, we found that not all of the most observable RBD mutations strengthen  viral infectivity  \cite{wang2021mechanisms}. This exception took place in the middle and late 2021 when a good portion of the population in many developed countries was vaccinated. By the genotyping of 2,298,349 complete SARS-CoV-2 genomes, we discovered vaccination-induced antibody-resistant mutations, which make the virus less infectious \cite{wang2021mechanisms}. This discovery leads to a complementary mechanism of natural selection, namely antibody-resistant mutations. In other words, viral evolution also favors RBD mutations in a population that enable the virus to escape antibody protection generated from   vaccination or infection.

The Omicron variant was the first example that was induced by both infectivity strengthening and antibody resistance mechanisms \cite{wang2021mechanisms}.  It has 32  mutations on the  S  protein, the main antigenic target of antibodies \cite{cele2021omicron}.  Among them, 15 are on the Omicron RBD, leading to a dramatic increase in SARS-CoV-2 infectivity, vaccine breakthrough, and antibody resistance \cite{chen2022omicron}. The World Health Organization (WHO) declared Omicron as a variant of concern (VOC) on November 26, 2021. On December 1, 2021, when there were no experimental data available, we announced our topological artificial intelligence  (AI) predictions based on the genotyping of viral genomes, biophysics, experimental data of protein-protein interactions, algebraic topology, and deep learning  \cite{chen2022omicron}. We predicted that Omicron is about 2.8 times as infectious as the Delta and has nearly 90\% likelihood to escape vaccines, which would compromise essentially all of existing monoclonal antibody  (mAb) therapies from Eli Lilly, Regeneron, AstraZeneca,  etc.    These predictions were subsequently confirmed by   experiments  \cite{shuai2022attenuated,cele2021omicron,zhang2022significant, liu2021striking,lu2021neutralization,hoffmann2021omicron}. On February 10, our topological AI model foretold the taking over of  Omicron BA.1 by Omicron subvariant BA.2 \cite{chen2022omicron2}. The WHO declared BA.2's dominance on March 26, 2022.  On May 1, 2022, our topological AI model projected the incoming dominance of BA.4 and BA.5 \cite{chen2022persistent}, which became reality in late June 2022. Currently,  BA.5 is still the world's dominant variant. Therefore, our topological AI model has been offering unusually accurate two-month forecasts of emerging dominant variants.   

The COVID fatigue and the worldwide relaxation of COVID-19 prevention measures have given the virus enormous new opportunities to spread in world populations, which enables the virus to further evolve. Additionally, the newly generated Omicron subvariant RBD structures leave abundant room for the virus to further optimize its binding with the ACE2 and disrupt existing antibodies, resulting in a large number of emerging subvariants.  It is of paramount importance to analyze their growth potentials in the world's populations and alert future dominant variants.    

This work analyzes SARS-CoV-2 evolutionary trends. We predict the  SARS-CoV-2 S protein  RBD mutation-induced binding free energy (BFE) changes of  RBD-human ACE2 complexes at all RBD residues. Such changes are employed to forecast Omicron subvariants' growth potentials and chances of becoming future dominant variants. Topological AI models are built from newly available deep mutational screening data and Omicron BA.1 and BA.2 three-dimensional (3D) structures. Our studies are assisted with the genotyping of over three million SARS-CoV-2 genomes extracted from patients and the evolutionary pattern of viral lineages among infections in the United States.

\section{Results}

We carry out single nucleotide polymorphism (SNP) calling for 3,616,783 million complete genomes extracted from patients. All unique mutations and their observed frequencies are illustrated in Fig. \ref{fig:mutationtracker}. Our interactive website, Mutation Tracker, also provides detailed records of mutations for download. On average, each nucleic acid site has one mutation.  Overall, mutations occur essentially randomly at all 29,903 bases. Therefore, simple SNP calling and genotyping does not offer any direct evidence for SARS-CoV-2 variants as discussed earlier. More specific analysis of the RBD mutations is used for the forecasting of future dominant variants.

\begin{figure}[ht!]
	\centering
	\includegraphics[width = 0.7\textwidth]{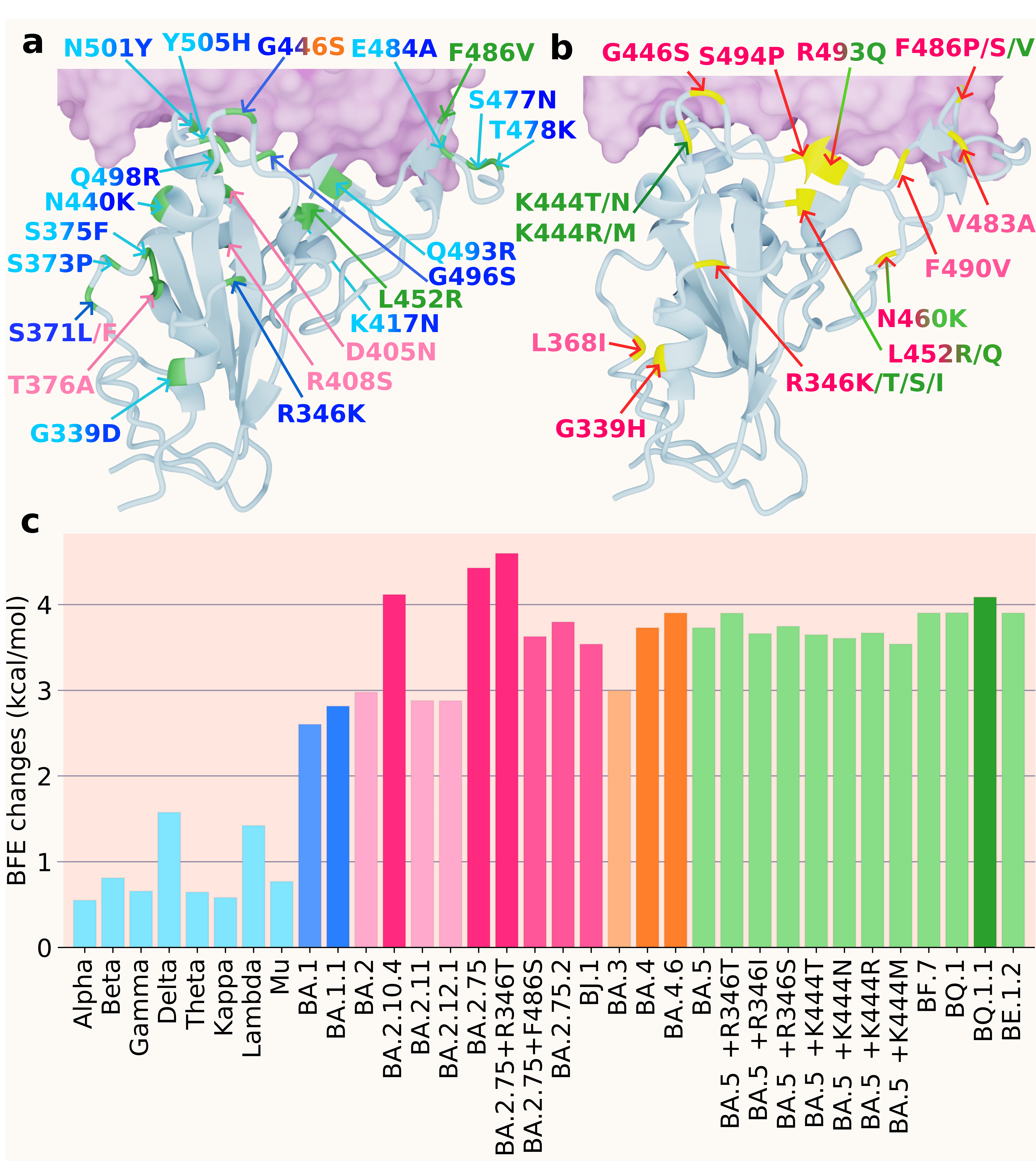}
	\caption{{\bf a}. and {\bf b}. the 3D structure of BA.2 (PDB: 7XB0~\cite{li2022structural}) with two sets of mutations (colors are consistent with those in {\bf c} and integrated colors indicate that mutation appears on multiple variants). {\bf a}. the mutations of precious VOCs (cyan) and BA.1 (blue), BA.2 (pink), BA.3 (orange)
, BA.4, and BA.5 (green). {\bf b}. the mutations of the Omicron subvariants   with BA.2 sublineages (pink) and BA.5 sublineages (green).   {\bf c}. A comparison of predicted mutation-induced BFE changes for previous VOCs and Omicron subvariants. Previous VOCs (cyan): Alpha, Beta, Gamma, Delta, Theta, Kappa, Lambda, Mu; BA.1 and BA.1.1 (blue); BA.2 and sublineages (pink): BA.2, BA.2.10.4, BA.2.11, BA.2.12.1, BA.2.75, BA.2.75+R346T, BA.2.75+F486S, BA.2.75.2, BJ.1 (BA.2.10.1.1); BA.3 (light orange); BA.4 and BA.4.6 (orange)
; BA.5 and sublineages (green): BA.5, BA.5+R346T/I/S, BA.5+K444T/N/R/M, BF.7 (BA.5.2.1.7), BQ.1 (BA.5.3.1.1.1.1.1), BQ.1.1 (BA.5.3.1.1.1.1.1.1), BE.1.2 (BA.5.3.1.1.2).
	}
	\label{fig:mutations}
\end{figure}

We collect emerging Omicron sublineages and compare them with previous VOCs.
We  use the notation ``*'' to represent the lineage and its sublineage. For instance, BA.2* represents BA.2 and all its sublineages in Figure ~\ref{fig:mutations}. Figure~\ref{fig:mutations}{\bf a} and {\bf b} show the 3D structures of RBD binding to human ACE2. Figure~\ref{fig:mutations}{\bf a} includes the RBD mutations of previous VOCs and Omicron subvariants  BA.1, BA.2, BA.3, BA.4, and BA.5, while Figure~\ref{fig:mutations}{\bf b} shows mutations   of the subvariants of Omicron BA.2 and BA.5. Lineages originated from BA.2  are marked in red type of colors. Subvariants originated from BA.5 are labeled in green type of colors. 
In Figure~\ref{fig:mutations}{\bf c}, the BFE changes of previous VOCs, BA.1 and BA.2 are calculated as the accumulation  of single mutations according to the original structure (PDB: 6M0J~\cite{lan2020structure}). The  BFE changes of BA.1.1 is calculated based on the BA.1 RBD-ACE2 structure (PDB: 7T9L~\cite{mannar2022sars}).
For the sublineages of BA.2, as well as BA.3, BA.4* and BA.5* (including BF.7, BQ.1*, and BE.1.2), their BFE changes are calculated based on the BA.2 structure (PDB: 7XB0~\cite{li2022structural}).

In Figure~\ref{fig:mutations}{\bf c}, the variants prior to the Omicron are presented in light blue. There are two main clades, one from   BA.2 and  the other  from   BA.5. Others, including previous VOCs, BA.1, BA.3, and BA.4, as well as their sublineages,
are also presented. Firstly,  three mutations from BA.2 to BA.5 are   L452R, F486V, and R493Q, which make BA.5  two-fold as infectious as BA.2.
This   explains why BA.5 replaced BA.2 as a new dominant variant in late June 2022. Among the BA.2 sublineages, BA.2.10.4, BA.2.75, and BA.2.75+R346T were predicted to 
have BFE changes greater than 4.0 kcal/mol.
These three sublineages together with  BA.2.10.4 and BA.2.75.2 
have higher BFE changes and are more infectious than BA.4 and BA.5.
As for BA.4 and BA.5 sublineages,
BA.4.6 is more infectious than  BA.4 and BA.5 and has   potential to become a dominant  variant.
Among the sublineages of BA.5,
BQ.1.1 has the highest potential to replace the spreading of BA.5 as its BFE change is greater   than 4.0 kcal/mol,
while BA.5+R346T, BF.7, BQ.1, and BE.1.2 have larger BFE changes than that of BA.5.
Based on this analysis, we forecast that BA.2.10.4, BA.2.75,
BA.2.75+R346T, 
%BA.4.6, BF.7 
and BQ.1.1 have high potentials to become new  dominant  variants.

\begin{figure}[ht!]
	\centering
	\includegraphics[width = 1\textwidth]{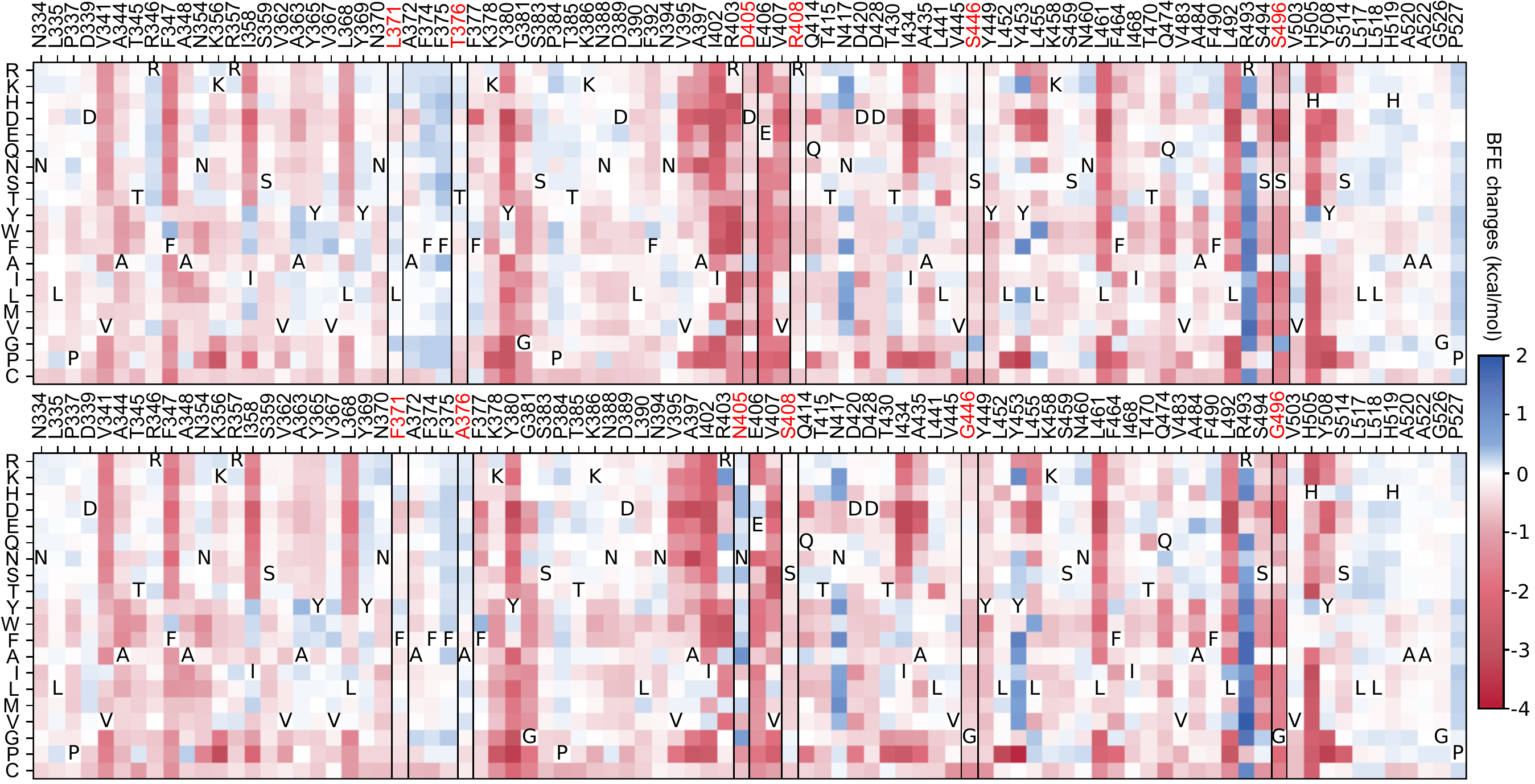}
	\caption{Heatmap of mutation-induced BFE change predictions of BA.1 (top panel) and BA.2 (bottom panel). Residues that have at least one mutation-induced BFE change greater than 0.1 kcal/mol are selected. The sites of BA.2's six distinct mutations  are marked red and framed in the heatmap.
	}
	\label{fig:heatmap}
\end{figure}
Figure~\ref{fig:heatmap} shows the heatmap of mutation-induced BFE change predictions of BA.1 (top panel) and BA.2 (bottom panel). 
We plot those residues that have at least one mutation-induced BFE change greater than 0.1 kcal/mol, which gives rise to 89 residues in the plots. In other words,  we keep mutations that will lead to more infectious variants. The deep blue color indicates infectivity-strengthening  mutations. Deep red color shows   infectivity-weakening  mutations.  
It is seen from  Figure~\ref{fig:heatmap} that most mutations will weaken the binding between RBD and ACE2 for BA.1 and BA.2. However, such mutations, once occurred, will have little chance  of becoming clinically significant due the natural selection.   Figure~\ref{fig:heatmap} indicates that both BA.1 and BA.2 are highly infectivity-optimized variants. They just leave a few residues  to be further optimized. Obviously, for both B.1 and BA.2, many mutations on residue sites R439, Y453, and N417 will most likely lead to more infectious new variants.  For BA.2, surveillance is also required for residue sites N504 and R403.    

  Compared with BA.2, BA.5 has three additional mutations, i.e., L452R,	F486V, and	R493Q. Among them, R493Q makes BA.5 significantly more infectious as shown  in Figure~\ref{fig:heatmap}. 
This reverse mutation (original residue is glutamine) occurs in many other lineages
showing in Figure~\ref{fig:mutations}{\bf c}, namely,
BA.2.10.4, BA.2.75*, BA.4*, BA.5, BF.7, BQ.1* and BE.1.2.
In addition to R493Q,
BA.2.75* and BQ.1.1 in Figure~\ref{fig:mutations} share the mutation N460K with the BFE change 0.267 kcal/mol.
This  indicates that more infectious variants will emerge 
with multiple infectivity-strengthening  mutations.
Overall, comparing the two heatmaps in Figure~\ref{fig:heatmap},
it is easy to  note that BA.2 has more positive BFE changes,
which makes future  BA.2 sublineages more competitive than future BA.1 sublineages in terms of infectivity. 

The top panel  of Figure~\ref{fig:heatmap}   explains why BA.2 is more infectious than BA.1.  BA.2 shares  12 of its RBD mutations with BA.1, except for six mutations, i.e., L371F, T376A, D405N, R408S, S446G, and S496G. These residue sites are marked with red in both panels of Figure~\ref{fig:heatmap}.  
%Interestingly, there is a negative correlation of the BFE changes between the mutations of BA.1 and BA.2.
Among these mutations,  L371F, T376A, D405N, and R408S induced minor BFE changes as shown in the top panel of Figure~\ref{fig:heatmap}. However, S446G  and S496G
 render  BA.2 significantly more infectious than BA.1. 
%Moreover, R493 on both BA.1 and BA.2 has many positive BFE changes and R493Q is one of them to increase the binding affinity.

\begin{figure}[ht!]
	\centering
	\includegraphics[width = 0.85\textwidth]{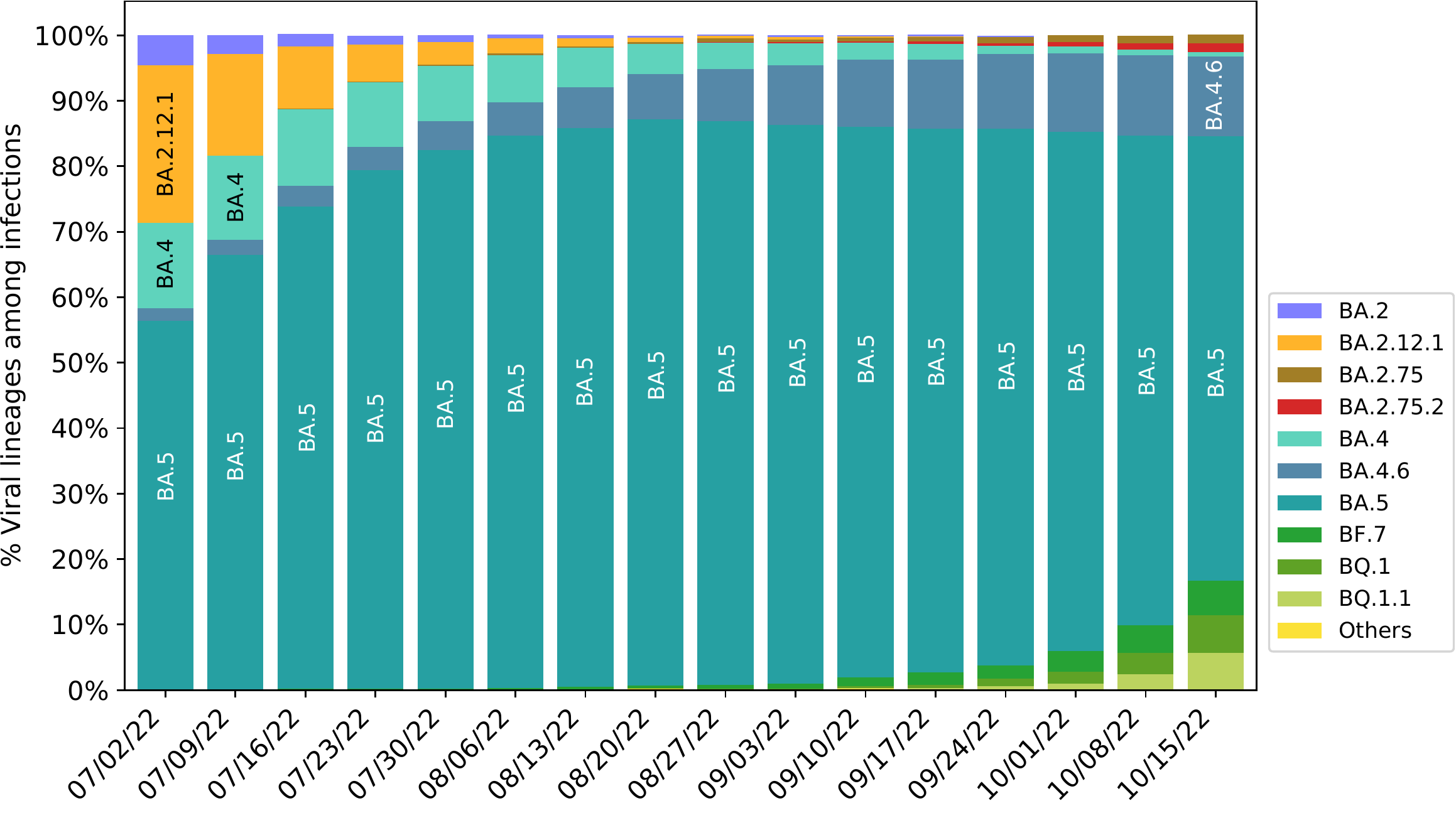}
	\caption{Weekly viral lineages among infections in the United States from 06/26/2022 to 10/08/2022. AY.1-AY.133, Delta (B.1.617.2), BA.1 and sublineages of BA.1 variant are aggregated to   category ``Others''. BA.2 sublineages except BA.2.12.1, BA.2.75 and BA.2.75.2, are aggregated with BA.2. BA.4 sublineages are aggregated to BA.4 except BA.4.6. Sublineages of BA.5 are aggregated to BA.5 except BF.7, BQ.1 and BQ.1.1. The spike substitution R346T is included in lineages BA.4.6, BF.7, and BA.2.75 Data from CDC website \cite{CDCVariantProportions}.
	}
	\label{fig:frequency}
\end{figure}

\section{Discussion}

Figure~\ref{fig:frequency} presents the evolution pattern of weekly viral lineage distribution among infections in the United States from 06/26/2022 to 10/08/2022. Each lineage is illustrated by  aggregating its sublineages to except for its sublineage is also listed. 
Note that BA.2.75 sublineages are aggregated to BA.2.75, 
which means  lineages BA.2.75+R346T and BA.2.75+F486S in Figure~\ref{fig:mutations} belong to this category.
It is interesting to note that there is high consistence between Figures \ref{fig:mutations} and ~\ref{fig:frequency}. Specifically, all the emerging variants listed in Figure~\ref{fig:frequency} have relatively high BFE changes as depicted in Figure \ref{fig:mutations}. 

It is also interesting to note from  Figure~\ref{fig:frequency}  that the relative populations  of BA.2.12.1, BA.4, and BA.5 are shrinking during this period.
BA.5 slightly expanded at the beginning and took a portion of BA.2.12.1's population.
  BA.4.6 is a sublineage of BA.4,
while BF.7, BQ.1, and BQ.1.1 are the sublineages of BA.5.
Their relative populations are increasing.
BQ.1.1 has a faster growth rate than BQ.1 and BF.7,
which indicates that the predicted  BFE change of BQ.1.1 is the highest among the sublineages of BA.5.
As shown in Figure~\ref{fig:mutations},
BA.2.75, 
%BA.4.6, BF.7, BQ.1, 
and BQ.1.1 have higher potentials to become future dominant variants. 

In our earlier predictions of Omicron \cite{chen2022omicron} and BA.2  \cite{chen2022omicron2}, we utilized nearly 200 antibody-RBD complexes to analyze the impact of antibody resistance. Such analysis is necessary because the Omicron variant involves a dramatic increase in the number of  RBD mutations. For the most variants studied in the present work, there are only gradual changes in the number of new RBD mutations and thus the impact of antibody resistance  on   natural selection may be  relatively small.      

 While the BFE change-based prediction favors the variant with the highest BFE change, its dominance in the population is also determined by the viral transmission environment (i.e., vaccination, prevention measures, human interaction intensity, etc.) and temporal dynamics. Therefore, a variant with slightly lower BFE change might become a dominant variant over a short period, which is called kinetic reaction control in thermodynamics. In an idealized viral transmission environment, the variant with the highest BFE change would have an exponential advantage over other variants, according to the Boltzmann distribution, which is called thermodynamic reaction control.

\section{Methods}

\subsection{Deep learning model  }

 The model applied in this work is an updated version of the recently proposed machine learning model, TopLapNet, by integrating the SKEMPI 2.0 dataset~\cite{jankauskaite2019skempi} and deep mutational scanning datasets~\cite{chan2020engineering,starr2020deep,linsky2020novo,starr2022deep}.
Briefly speaking, the TopLapNet model is a deep neural network model
and implements biophysics and biochemistry descriptors, as well as mathematical descriptors based on algebraic topology \cite{zomorodian2005computing,edelsbrunner2008persistent,wang2020persistent}
to predict the binding free energy (BFE) changes of protein-protein interactions (PPIs) 
induced by single mutations.
A deep neural network maps sample features to an output layer
where hidden layers in the network contain numerous neuron units and weights
updated by backpropagation methods.
The single neuron gets fully connected with the neurons
in the following layers.
For the model cross-validations, the Pearson correlation of 10-fold cross-validation is 0.864, and the root mean square error is 1.019 kcal/mol. 
As for predictions,
the TopLapNet model is used to calculate all possible mutation impacts on RBD binding to ACE2 
for the original virus (PDB: 6M0J~\cite{lan2020structure}), BA.1 (PDB: 7T9L~\cite{mannar2022sars}), and BA.2 structures (PDB: 7XB0~\cite{li2022structural}).
Thus, previous VOCs' infectivities as well as that of BA.1 and BA.2 are calculated based on the original structure. The infectivity of BA.1.1 is calculated by accumulating BFE changes based on the BA.1 structure. The infectivities of all other sublineages presented in Figure~\ref{fig:mutations} are calculated by the accumulations of BFE changes based on the BA.2 structure.

\subsection{Feature generation }
Feature generation methods decipher protein structures 
to extract their biophysics, biochemistry, and mathematical information.
These methods use physical, chemical, and mathematical modeling
of protein structures to provide suitable features for machine-learning algorithms.
There are two types of features, i.e., residue-level ones and atom-level ones.
Residue-level features are generated from secondary structures, 
which are provided by a position-specific scoring matrix (PSSM) 
in the form of conservation scores of each amino acid~\cite{altschul1997gapped}.
Atom-level features consider seven groups of atom types, 
including C, N, O, S, H, all heavy atoms, and all atoms.
Surface areas, partial changes, atomic pairwise interactions, and electrostatics
are assembled in an element-specific manner in terms of these seven groups.
Moreover, the most important features from modelings are topological features and graph features generated by using persistent homology \cite{zomorodian2005computing,edelsbrunner2008persistent } and persistent  Laplacian \cite{wang2020persistent}.

Persistent homology describes proteins by analogy to point cloud data.
Atoms are regarded as vertices to build a simplicial complex, 
which is a collection of infinitely many simplicies such as nodes, edges, triangles, and tetrahedrons.
The simplicies  among atoms are defined by
whether there is an overlap under a given influence domain or radius $r$.
Filtration of this topological space is defined by varying the radius
as a sequence of snapshots of each simplicial complex to extract more geometric and topological properties.
Then, the Betti numbers on each snapshot are computed as descriptors
of the number of connected components, cycles, and cavities in a protein structure.
Persistent  Laplacian (also known as persistent spectral graph \cite{ wang2020persistent})  on the other hand
unveil the  homotopic shape evolution of a protein structure in filtration 
that the persistent homology cannot provide.
Persistent   Laplacian applies the same scheme as persistent homology
to construct simplicial complexes during filtration.
However, persistent  Laplacian calculates all eigenvalues
of the combinatorial Laplacian with boundary operators on simplices.
Our mathematical features consist of both topological invariants from persistent homology and spectral invariants from persistent Laplacian.

\subsection{SNP calling and Mutation Tracker}
For genotyping,  SARS-CoV-2 complete genome sequences with high coverage and exact collection date were downloaded from the GISAID database \cite{shu2017gisaid} (\url{https://www.gisaid.org/}) as of September 30, 2022. Such sequences were aligned to the reference genome downloaded from GenBank (NC\_045512.2)\cite{wu2020new}. Next, we applied single nucleotide polymorphism (SNP) calling \cite{yin2020genotyping,kim2007snp} to measure the genetic variations between SARS-CoV-2 sequences through Cluster Omega with default parameters.
The SNP calling can track differences between various SARS-CoV-2 sequences and the reference genome. By applying it, we decoded 29,290 unique single mutations from more than 3.6 million complete SARS-CoV-2 genomes. The detailed mutation information can be viewed at \href{https://users.math.msu.edu/users/weig/SARS-CoV-2_Mutation_Tracker.html}{Mutation Tracker}.
Lastly, the Omicron sublineages analyzed in Figure~\ref{fig:mutations} are selected from the SNP analysis and other web-servers\cite{CDCVariantProportions,NIHvarianttrack,covSPECTRUM}. 

%\section{Conclusion}

\section*{Supporting information}
The Supporting information (Supporting\_Material.zip) includes
the predicted BFE changes for all possible RBD mutations based on BA.1 and BA.2 structures, PDB files (6M0J, 7T9L, 7XB0) and a README file for the links of external datasets. 

\section*{Data and model availability}

The world’s SARS-CoV-2 SNP data are available at Mutation
Tracker (https://users.math.msu.edu/users/weig/SARS-CoV-2\_Mutation\_Tracker.html).   The TopNetTree
model is available at  https://github.com/WeilabMSU/TopNetmAb.

\section*{Acknowledgment}
This work was supported in part by NIH grants  R01GM126189 and  R01AI164266, NSF grants DMS-2052983,  DMS-1761320, and IIS-1900473,  NASA grant 80NSSC21M0023,  MSU Foundation,  Bristol-Myers Squibb 65109, and Pfizer.

\end{document}